\documentclass[]{aa}
\usepackage{graphicx}
\usepackage{txfonts}

\begin{document}

\title{Observation of kink waves in solar spicules}

\author{V. Kukhianidze\inst{1}, T.V. Zaqarashvili\inst{1,2} \and E. Khutsishvili\inst{1}}

\institute{Abastumani Astrophysical Observatory, Kazbegi Ave. 2a, Tbilisi 0160, Georgia, \\
\email{[vaso;temury;eldar]@genao.org} \and Departament de F\'{\i}sica,
Universitat de les Illes Balears, E-07122 Palma de Mallorca, Spain, \\ \email{temury.zaqarashvili@uib.es}}

\offprints{T.V. Zaqarashvili}

\date{Received / Accepted }

\abstract{Height series of H${\alpha}$ spectra in solar limb
spicules obtained with the 53 cm coronagraph of the Abastumani
Astrophysical Observatory are analyzed. Each height series covered 8
different heights beginning at 3800 km above the photosphere. The
spatial difference between neighboring heights was 1$^{\prime
\prime}$, consequently $\sim$ 3800 - 8700 km distance above the
photosphere has been covered. The total time duration of each height
series was 7 s. We found that nearly 20$\%$ of measured height
series show a periodic spatial distribution of Doppler velocities.
We suggest that this spatial periodicity in Doppler velocity is
caused by propagating kink waves in spicules. The wave length is
found to be $\sim$ 3500 km. However the wave length tends to be
$\sim$ 1000 km at the photosphere due to the height variation of
the kink speed. This probably indicates to a granular origin for the
waves. The period of waves is estimated to be in the range of 35-70
s. These waves may carry photospheric energy into the corona,
therefore can be of importance in coronal heating.

}

\maketitle

\section{Introduction}

The heating of the upper chromosphere and corona is still an unsolved
problem in solar physics. It is clear that the energy source
supporting the high temperature lays in the highly dynamical and
dense photosphere. There the mechanical energy of photospheric
motions can be guided upwards by structured magnetic fields in the
form of waves or electric currents. Therefore the energy transport
by magnetohydrodynamic (MHD) waves throughout the solar atmosphere
is one of the key process towards the solution of the heating
problem (Roberts \cite{rob1}). Observations of oscillatory phenomena
in the solar atmosphere have increased dramatically in the last few
years through observations from SOHO (Solar and Heliospheric
Observatory) and TRACE (Transition Region and Coronal Explorer). The
space missions uncovered the rich spectrum of MHD oscillations in
the transition region and corona (Doyle et al. \cite{doy}; 
Nakariakov et al. \cite{nak};  Ofman et al. \cite{ofm}; Banerjee et al. \cite{ban}; O'Shea
et al. \cite{osh}; De Moortel et al. \cite{moor}). The coronal waves are either
generated in situ or they penetrate from the photosphere. Hence the
observation of oscillatory phenomena in chromospheric spectral lines
is of vital importance.

Most of the chromospheric radiation in quiet Sun regions comes from
spicules, which are jet-like chromospheric structures observed at
the solar limb mainly in H${\alpha}$ line. Spicules often show the
group behaviour and probably are concentrated between supergranule
cells (see e.g. reviews of Beckers \cite{bec} and Sterling
\cite{ste}). Therefore spicules probably are formed in regions of
magnetic field concentration and consequently MHD wave propagation
in the solar atmosphere may be traced through their dynamics.
Oscillations in spicules with $\sim$5 minute period have been
detected by ground based (Kulidzanishvili \& Zhugzhda \cite{kul})
and recently by space observations (De Pontieu et al. \cite{dep};
Xia et al. \cite{xia}). On the-other-hand, oscillations in spicules
with shorter period have been reported more than 30 years ago by
Nikolsky \& Platova (\cite{nik}). They found that spicules oscillate
along the limb with a characteristic period of about 1 min. The
oscillation, which they reported, was a periodic transversal
displacement of the spicule axis at one particular height.

If spicules are formed in thin magnetic flux tubes, then the
periodic transverse displacement of the axis observed by Nikolsky \&
Platova (\cite{nik}) probably was due to the propagation of kink
waves. It is well known that transverse kink waves can be generated
in photospheric magnetic tubes by buffeting of granular motions
(Roberts \cite{rob2}; Hollweg \cite{hol}; Spruit \cite{spru}; Hasan
\& Kalkofen \cite{has}). Kink waves cause the displacement of the
tube axis, therefore their propagation can be traced either by
direct observation of the tube displacement along the limb as in
Nikolsky \& Platova (\cite{nik}) or spectroscopically by the Doppler
shift of spectral lines. The later possibility arises when the
velocity of kink waves is polarized in the plane of observation. The
periodic spatial distribution of Doppler velocities in spicules has
been detected almost 20 years ago by Khutsishvili (\cite{khu}).
Unfortunately, neither observers nor theorists paid attention to the
phenomenon which can be simply explained in terms of kink waves. In
this letter, we reanalyze the height series of H${\alpha}$ spectra
obtained by Khutsishvili (\cite{khu}) and then model the observation
as propagating kink waves.

\begin{figure}
\centering
\includegraphics[width=0.8\linewidth]{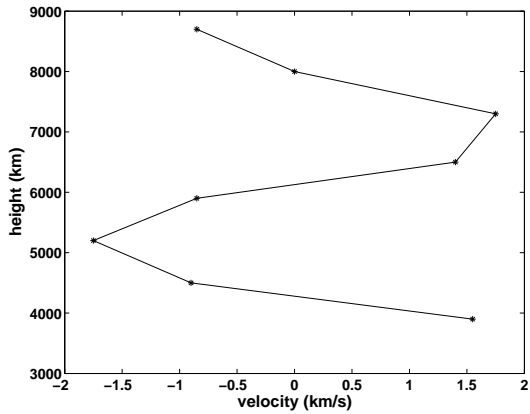}
\caption{The Doppler velocity spatial distributions for one of the
height series is shown. Marked dots indicate the observed heights.
The Doppler velocity has clear periodic spatial distribution, which
indicates wave propagation. \label{fig:single}}
\end{figure}

\section[]{Observation}

The big (53 cm) coronagraph and universal spectrograph of Abastumani
Astrophysical Observatory has been used to obtain chromospheric
H${\alpha}$ line spectra at different heights (8 heights) above the
photosphere (Khutsishvili \cite{khu}). Instrumental spectral
resolution is 0.04 {\AA} and dispersion 1 {\AA}/mm in H${\alpha}$.
Observation has been carried out at the solar limb as height series
beginning at 3800 km above the photosphere with a step size
of 1 arcsec. Thus the distance $\sim$ 3800-8700 km above the
photosphere was covered. The exposure time was 0.4 s at lower
heights and 0.8 s at higher ones. The duration of each height series
was 7 s, with the total duration of the observation being 44 min.
More details about the observation can be found in Khutsishvili
(\cite{khu}).

We analyzed spatial distributions of Doppler velocities in selected
H${\alpha}$ height series. Nearly 20$\%$ of measured height series
showed a periodic spatial distributions in the Doppler velocities.
The Doppler velocity spatial distributions for one of the height series
is shown in Fig.1. Periodic spatial distribution is clearly seen,
which indicates a wave propagation. The wave length can be estimated
as ${\sim}$ 3500 km. The maximal time difference between the
observations at the lowest and highest heights in one series is 7 s. If
the wave propagates with the Alfv{\'e}n speed, being say $\sim$ 50
km/s in the chromosphere, then during 7 s it will pass a distance
$\sim$ 350 km. So during one height series the wave will propagate
along the distance which is less than the distance between
consequent observed heights. Therefore the spatial structure of the
wave velocity in one height series can be considered as nearly
simultaneous.

\begin{figure}
\centering
\includegraphics[width=0.7\linewidth]{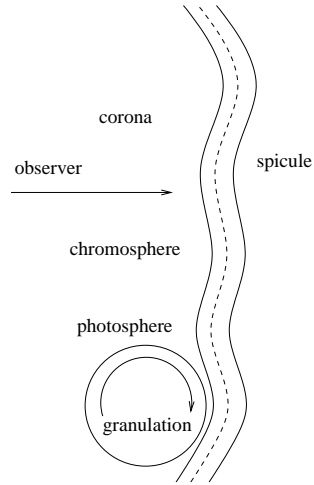}
\caption{Schematic picture of propagating kink waves in spicules.
Due to the kink waves, the observed spectral line is Doppler shifted
when the wave velocity is polarized in the plane of observation.
\label{fig:single}}
\end{figure}

Thus the observations show the evidence of transversal wave
propagation in spicules with typical wave length of ${\sim}$ 3500
km. The waves can be either kink or linearly polarised Alfv{\'e}n
waves (torsional Alfv{\'e}n waves in magnetic tubes will cause
periodic broadening of spectral lines not the Doppler shift, see
Zaqarashvili \cite{zaq}). But as spicules are highly structured
phenomena (their width are $\sim$ 1$^{\prime \prime}$), the observed
Doppler shift can be caused due to the propagation of kink waves.

Kink waves propagating along the magnetic tube lead to the
oscillation of the tube's axis (see Fig.2). Therefore if the
velocity of the kink wave is polarized in the plane of observation,
then it results in the Doppler shift of the observed spectral line.
The Doppler shift will have a periodic behaviour in height: at the
same time, the spectral line will have blue shift at antinodes where
the tube moves towards the observer and red shift where the tube
moves in the opposite direction; at the velocity nodes the Doppler
shift tends to zero. This means that kink waves manifest itself in
the same spatial behaviour of Doppler shift as revealed by our
height series (Fig.1). However if the spicule axis is inclined with respect
to the local vertical direction then a steady flow inside the spicule will shift
the Doppler velocity at all heights with the same value. Indeed,
some spicules show this behaviour. Three consecutive height series
of Doppler velocity in one spicule are shown in Fig.3. The velocity
is shifted by $\sim$ 12 km/s at all heights, although the wave
signature is still seen. If the spicule is inclined by approximately $35\ ^{\circ}$ 
(Trujillo Bueno et al. \cite{tru}) then the real steady flow velocity will be $\sim$ 22
km/s, similar to the typical mass raising speed in spicules. Also
Fig.3 shows that the maximum of the Doppler velocity moves up in
consecutive height series. This may indicate a wave phase
propagation. The phase is displaced at $\sim$ 1500 km in about 18 s
giving the phase speed of $\sim$ 80 km/s, very similar to the expected
kink speed at these heights.

\begin{figure}
\centering
\includegraphics[width=0.9\linewidth]{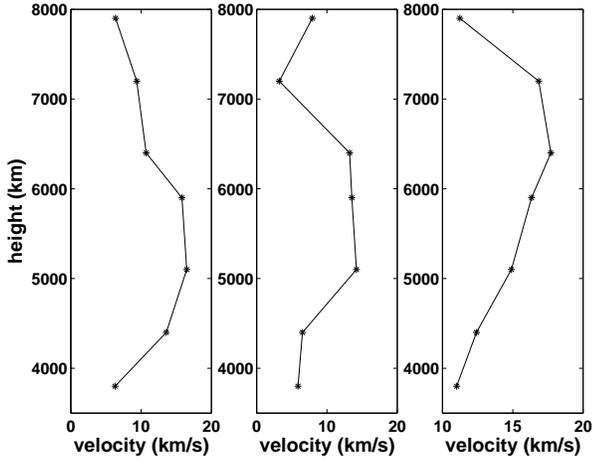}
\caption{Three consecutive height series of Doppler velocity in one
spicule. The time difference between the consecutive plots is $\sim$ 8 s.
The maximum of Doppler velocity moves up in consecutive height series, which probably indicates a wave propagation. \label{fig:single}}
\end{figure}

\section{The model}

The mechanism of spicule formation is still not well understood (see
the recent review of Sterling \cite{ste} and references therein).
Therefore this letter does not address the spicule formation
mechanism. Spicule life time is $\sim$ 10 -15 min, while the period
of observed waves is much shorter: 70 s for the wave with phase
speed of 50 km/s and wave length of 3500 km. In 2-3 periods the wave
will propagate along the whole spicule length. During this short
time intervals spicules can be considered as existing stable
structures. Therefore the waves may propagate independently of the
spicule formation mechanism.

During short time intervals spicules can be modeled as thin magnetic
flux tubes embedded in a field free environment, anchored in the
photosphere and persisted towards the corona. Propagation of kink
waves in vertical magnetic tube embedded in a field free environment
is governed by the equation (Roberts \cite{rob1})
\begin{equation}
{{\partial^2 {\xi}}\over {\partial t^2}}= c^2_k{{\partial^2
{\xi}}\over {\partial z^2}} + g{{\rho_0 - \rho_e}\over {\rho_0 +
\rho_e}}{{\partial {\xi}}\over {\partial z}},
\end{equation}
where $\xi$ is the transverse displacement, $\rho_0(z),\, \rho_e(z)$
are the plasma densities inside and outside the tube accordingly,
$g$ is the gravitational acceleration and $c_k=c_A\sqrt{{{\rho_0}/
{\rho_0 + \rho_e}}}$ the kink speed. Here $c_A={B_0/{\sqrt
{4\pi\rho_0}}}$ is the Alfv{\'e}n speed with $B_0(z)$ as the tube
magnetic field. Note that this equation does not include the
flow along the tube, while spicules show continuous upward mass
motion. Therefore equation (1) may correctly describe the waves only 
in the co-moving frame (with the steady flow), and thus the wave phase speed will be
Doppler shifted due to the mass motion inside the spicule, which 
may slightly alter the theoretical results.

Spicule density and magnetic field show almost no spatial variation
at observation heights ($\sim$ 3800-8700 km above the photosphere)
and the stratification also can be neglected in this part of the
solar atmosphere, then equation (1) gives the approximate dispersion
relation $c^2_k k^2_z = \omega^2$, where $k_z$ is the vertical wave
number and $\omega$ is the frequency of kink waves. From the
dispersion relation we may estimate the period of kink waves using
the observed wave length and typical kink speed. The observed wave
length of kink waves is of order $\sim 3500$ km. Then
for the kink speed of $\sim 50-100$ km/s at heights of 3800-8700 km
above the photospheric, the period of kink waves can be
estimated as ${\sim}$ 35-70 s.

\begin{figure}
\centering
\includegraphics[width=0.9\linewidth]{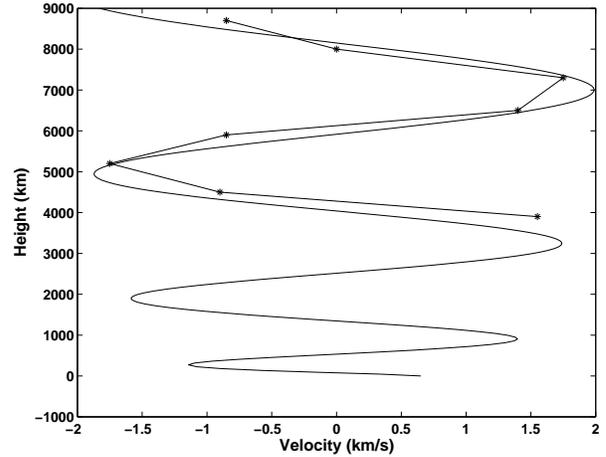}
\caption{Comparison of observed Doppler velocity plot (Fig.1) and
the numerical solution of the kink wave equation. Vertical axis
shows the height in km and the horizontal axis shows the kink wave
velocity in km/s. It is seen that the wave length of the computed
kink waves at the photospheric level is $\sim$ 1000 km, which is
comparable to the granular diameter. \label{fig:single}}
\end{figure}

It can be suggested that the source of observed kink waves resides
in the lower part of the solar atmosphere. The cut-off period of
kink waves due to stratification at the photosphere (for the
pressure scale height of 125 km and plasma $\beta$ of $\sim$ 1) is
$\sim$ 660 s. So the expected period of kink waves is well below the
cut-off value. Thus the kink waves with periods of $\sim$ 35-70
s may easily propagate upwards. As estimated frequency of the
observed waves is much higher than the frequency of 5-minute
oscillations, then the oscillations can be ruled out to be the
source of kink waves. Therefore, the only source which may excite
the kink waves in the photosphere is the granulation. The
photospheric granulation has been often suggested as the source of
kink waves in thin magnetic tubes (Roberts \cite{rob2}; Spruit
\cite{spru}; Hollweg \cite{hol}; Hasan \& Kalkofen \cite{has}).
Photospheric granulation is very dynamic even during the life time
of one granular cell. If the magnetic tube, in which the spicule is
formed, is anchored in the photosphere, then any perturbation of the
granular cell probably excites the kink waves with wave length
similar to the cell diameter (see Fig.2). But the observed wave
length ($\sim$ 3500 km) of the kink waves in the higher atmosphere is a
few times longer than the mean granular diameter, which is of order
$\sim$ 800 -1000 km. The discrepancy can be resolved by an increase
of wave phase speed with height. Indeed, the kink speed being $\sim$
10 km/s at the photosphere increases up to $\sim$ 50 km/s (or even
more) in the higher atmosphere. Therefore it can be suggested that
granular cells generate kink waves with a wave length comparable to
their diameter, but the wave length increases due to increasing
phase speed when waves propagate upwards.

In order to show this, one should solve the kink wave equation (1)
along the whole solar atmosphere from the photosphere to the present
observed heights. As equation (1) is homogeneous with time, we may
perform a Fourier expansion in time in the form
$\xi={\xi}_1(z)\exp({i\omega t})$, where $\omega$ is the wave
frequency. Then its numerical solution is straightforward; we only
need to fix the spatial variations of the unperturbed parameters.
Unfortunately, the height variation of the physical quantities is
not well known. There are only suggested values of density and
magnetic field in spicules and they are also controversial. Also the
difference between plasma densities inside and outside of spicule
are not well determined. At the photospheric level the thermal
equilibrium inside and outside the tube i.e. $T_0(z)=T_e(z)$ (here
$T_0(z),\,\,T_e(z)$ are plasma temperatures inside and outside of
tube) is a good approximation, which gives a higher density outside
the tube than inside (in order to hold the transverse pressure
balance at the tube boundaries). But in the higher atmosphere, say at
heights $>$ 2000 km, the plasma temperature is much lower in
spicules than in the environment. On the-other-hand, the plasma
density seems to be much higher in spicules than the surroundings.
Due to this uncertainty in the medium parameters, it is better to
give the height dependence of the Alfv{\'e}n speed and the density
ratio inside and outside the spicule $\rho_0/\rho_e$. Therefore we take
the height dependence of the Alfv{\'e}n speed as $10$ km/s at the
photosphere and $\sim$ 100 km/s at 6000 km. For the density ratio we
take 1/3 at the photosphere (being higher outside the tube), which
increases up to 10$^3$ at a height of 6000 km. With this spatial
dependence of the Alfv{\'e}n speed and the density ratio we solve
numerically the kink wave equation for different values of frequency
$\omega$. Then we fit the numerical solution to the observed curve
of the Doppler velocity. The numerical solution is best fitted to
the observation (Fig.4) when the wave period is $\sim$ 40 s. From
Fig. 4 it is seen that the wave length of the kink waves at the
photospheric level is $\sim$ 1000 km, which is comparable to the
granular diameter. It clearly indicates a granular origin for the
waves.

\section{Discussion}
Kink oscillations of coronal loops have been frequently observed in
the solar corona by TRACE as periodic loop displacement in space.
However in the lower atmosphere the observation of kink waves is
complicated due to their nearly incompressible character. Here we
report the spectroscopic observation of kink waves in solar limb
spicules. We have shown that nearly 20$\%$ of measured H${\alpha}$
height series at the solar limb ($\sim$ 3800 - 8700 km distance
above the photosphere) show the periodic spatial distribution of
Doppler velocities in spicules, which can be caused by propagating
kink waves. The wave length at these heights is $\sim$ 3500 km,
which goes to $\sim$ 1000 km at the photosphere if the height
dependence of the kink speed is taken into account. The estimated
wave period is $\sim$ 35-70 s. The computed wave length at the
photospheric level is comparable to the granular size, therefore the
granulation is probably the most plausible source for the wave
excitation. Then the granulation will excite kink waves in any thin
magnetic flux tube anchored in the photosphere. As we already noted
only $\sim$ 20$\%$ of measured height series show periodic spatial
distribution of Doppler velocities. Indeed spectroscopic detection
of kink waves is possible only when the wave velocity is polarised
in the plane of observation. But the granulation will excite the
waves with any direction of polarisation; there is no preferred
direction. Therefore 20$\%$ of measured height series is a good
percentage for the observation of waves.

It must be noted that some spicules show a multi-component
structure (i.e. several spicules are located close together) mostly
at lower heights, which may give the impression of a velocity shift
(Xia et al. \cite{xia}). Therefore we choose only the spicules with well defined single-component
structure. We also note that the data analysis has been made only
for selected height series. Now we are doing the data reduction of
the whole observation, which will be analysed in the near future.

Here we do not address the mechanism of wave excitation in detail as
it is beyond the scope of this letter. But we suggest that
granular cells will frequently excite the kink waves with the wave
length comparable to their diameter in anchored magnetic tubes. The
waves may propagate upwards carrying the photospheric energy into
the corona, thus may cause significant input into the coronal
heating. Therefore detailed spectroscopic search of waves with
period of $\sim$ 40-60 s and wave length of $\sim$ 3000-4000 km is
desirable to be performed in the future.

\section{Acknowledgements}
The work of T.Z. was partially supported by the NATO Reintegration
Grant FEL.RIG 980755 and MCyT grant AYA2003-00123. We would like to thank the
referee, J. Doyle, for his suggestions which improved this letter.


\begin{thebibliography}{}

\bibitem[2001]{ban}Banerjee, D., O'Shea, E., Doyle, J. G. and Goossens, M., 2001, A\&A, 380, L39

\bibitem[1972]{bec}Beckers, J.M., 1972, ARA\&A, 10, 73

\bibitem[2004]{moor}De Moortel, I., Hood, A. W., Gerrard, C. L. and Brooks, S. J., 2004, A\&A, 425, 741

\bibitem[2003]{dep}De Pontieu, B., Erd{\'e}lyi, R. and de Wijn, A.G., 2003, ApJ,
595, L63

\bibitem[1999]{doy}Doyle, J. G., van den Oord, G. H. J., O'Shea, E. and Banerjee, D., 1999, A\&A, 347, 335

\bibitem[1999]{has}Hasan, S.S. and Kalkofen, W., 1999, ApJ, 519,
899

\bibitem[1986]{khu} Khutsishvili, E., 1986, Solar Phys. 106, 75

\bibitem[1983]{kul}Kulidzanishvili, V.I. and Zhugzhda, Y.D., 1983, Solar Phys. 88,
35

\bibitem[1981]{hol}Hollweg, J.V., 1981, Solar Phys. 70, 25

\bibitem[1999]{nak}Nakariakov, V.M., Ofman, L., Deluca, E.E., Roberts, B. \& Davila, J.M. 1999, Science, 285, 862

\bibitem[1971]{nik}Nikolsky, G.M. and Platova, A.G., 1971, Solar
Phys. 18, 403

\bibitem[2000]{ofm}Ofman, L., Romoli, M., Poletto, G., Noci, G. and Kohl, J. L., ApJ, 529, 592

\bibitem[2001]{osh}O'Shea, E., Banerjee, D., Doyle, J. G., Fleck, B. and Murtagh, F., 2001,  A\&A, 368, 1095

\bibitem[1979]{rob2}Roberts, B. 1979, Solar Phys. 61, 23

\bibitem[1981]{rob0}Roberts, B. 1981, In J.H. Thomas and L.E. Cram (eds), {\it The
Physics of Sunspots}, Sunspot, Sacramento Peak Observatory, 369

\bibitem[2004]{rob1}
Roberts, B. 2004, In Proc. SOHO 13 `Waves, Oscillations and
Small-Scale Transient Events in the Solar Atmosphere: A Joint View
from SOHO and TRACE', Palma de Mallorca, Spain, (ESA SP-547), 1

\bibitem[1981]{spru}Spruit, H.C. 1981, A\&A, 98, 155

\bibitem[2000]{ste}Sterling, A.C., 2000, Solar Phys., 2000, 196,
79

\bibitem[2005]{tru}Trujillo Bueno, J., Merenda, L., Centeno, R., Collados, M., Landi Degl´Innocenti, E., 2005, ApJ, 619, L191

\bibitem[2005]{xia}Xia, L.D., Popescu, M.D., Doyle, J.G. and Giannikakis, 2005, A\&A, 438, 1115

\bibitem[2003]{zaq}Zaqarashvili, T.V. 2003, A\&A, 399, L15

\end{thebibliography}
\end{document}